\documentclass[aps,prb,10pt,twocolumn,letterpaper,superscriptaddress,floatfix,longbibliography]{revtex4-2}

\usepackage{amsmath}
\usepackage{amssymb}
\usepackage{amsfonts}
\usepackage{bm}
\usepackage{mathrsfs}
\usepackage{tensor}
\usepackage{dsfont}
\usepackage{esint}
\usepackage{comment}
\usepackage[caption=false]{subfig}

\usepackage{graphicx}
\usepackage{booktabs}

\usepackage[dvipsnames]{xcolor}
\usepackage{physics}
\usepackage[section]{placeins}

\usepackage[hyperindex,pdftex, breaklinks,colorlinks = true,linkcolor = blue,urlcolor=blue,citecolor=blue]{hyperref}

\begin{document}
	
	\title{Multi-photon schemes for mid-infrared detection
    : Comparative study of bulk GaAs and \texorpdfstring{Ge$_{1-x}$Sn$_x$}{GeSnx}}

        \author{Alistair H. Duff}
	\email{a.duff@utoronto.ca}
	\affiliation{Department of Physics, University of Toronto, Toronto, Ontario M5S 1A7, Canada}
 
	\author{J. E. Sipe}
	\email{john.sipe@utoronto.ca}
	\affiliation{Department of Physics, University of Toronto, Toronto, Ontario M5S 1A7, Canada}
	
	\date{\today }
	
	\begin{abstract}
             We calculate the theoretical non-degenerate two photon absorption and three color injected current response tensors for bulk GaAs and Ge$_{1-x}$Sn$_x$ for a range of alloy compositions. In particular, by including a ``pump" beam we compare two ``schemes" that are sensitive to mid-infrared photons. In ``scheme I" we consider GaAs and a pump photon with energy greater than half the band gap, and in ``scheme II" we consider Ge$_{1-x}$Sn$_x$ with a pump photon with energy less than half the band gap. We find that for certain pump and alloy concentrations Ge$_{1-x}$Sn$_x$ has a substantially larger nonlinear response and three-color injected current than GaAs in the mid-infrared frequency window where both materials can absorb photons via non-degenerate two-photon absorption.   
        \end{abstract}
	
	\maketitle

\section{Introduction}\label{Sec1}

Mid-infrared (MIR) photons -- those with wavelengths between 2.5 $\mu$m to 20 $\mu$m, the molecular ``fingerprint" region -- have a host of applications, ranging from chemical and biological sensing \cite{OnChipMIR,MIR_QO_Silicon} to free space optical communication \cite{MIR_FreeSpace}. Yet detection in this wavelength range is problematic: Semiconductor-based devices made of materials like InSb or HgCdTe alloys \cite{MIRSpectroscopy_Tutorial,InfraredDetectors}, superconducting nanowire single photon detectors \cite{MIRSpectroscopy_Tutorial,SNSPD_1,SNSPD_2}, transition edge sensors \cite{TES}, and microwave kinetic inductance detectors \cite{MKIDs} are available, but all require cryogenic temperatures to operate.

In addition to the linear absorption processes that underlie the operation of such devices, however, there are multi-photon absorption processes where the simultaneous absorption of $n$-photons leads to a single electronic transition \cite{Wherrett}. They could provide an alternate route to MIR detection: Photons which individually have energy less than the band gap can still be used to excite a transition, since it is only the sum of the photon energies that must cross the band gap. As a preliminary measure of detection efficiency at a given intensity, we can consider the number of carriers in the material promoted to the conduction bands, or the amount of energy absorbed from the electromagnetic field by the material. Both are described by the nonlinear absorption coefficient. In this manuscript, we focus on two-photon absorption (2PA), where one of the photons is provided by a strong pump beam. In general, higher order processes that involve even more photons are much weaker at pump intensities that do not damage the material \cite{NL_Optics}. 

\begin{figure}[ht]
    \centering
    \includegraphics[width=0.48\linewidth]{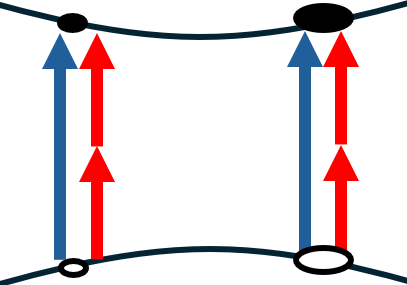}
    \caption{Schematic representation of the generation of current through the interference of degenerate 2PA (D-2PA) at the fundamental frequency $\omega$ and 1PA at the harmonic $2\omega$. The larger ovals on the right indicate constructive interference for the electron-hole transition at positive $k$, versus destructive interference at $-k$. }
    \label{fig.Injection_Schematic}
\end{figure}

When investigation is extended to 2PA processes, there are other concomitant physical processes that also arise as possibilities for detection. Under illumination by multi-chromatic light,  the interference of different absorption pathways with the same initial and final state, for example 1PA and 2PA occurring simultaneously, can ``coherently drive a current \cite{PhysRevLett.76.1703,OpticalInjectionSipe,CoherentControlExperiment,CoherentControlBulkGaAs,VectorizedOptoElectronic,DirectionalPhotoCurrentQuIC,QuICStructuredLight,PhysRevB.60.R11257,CoherenceControlPhysica};" see Fig. \ref{fig.Injection_Schematic}. Typically, these studies have involved two-color light sources at different harmonics to drive distinct $m$- and $n$-PA processes, with the effect measured in materials such as 
bulk GaAs \cite{CoherentControlBulkGaAs}, AlGaAs \cite{CoherentControlExperiment}, and the topological insulator Bi$_2$Se$_3$ \cite{CoherentControlTopoInsulator}. In two-color schemes, the two photons in the 2PA process are degenerate (D-2PA), and so $\hbar\omega$ must be greater than half the band gap $E_\text{gap}$ for D-2PA to occur.

However, it has been shown both theoretically and experimentally that non-degenerate (ND) 2PA -- where the energy of the two-photons can be drastically different -- can provide significant enhancement over D-2PA \cite{BandGapDependence_2PA,BandGapDependence_ND2PA,ND_Gain_GaAs,ND2PA_QuantumWells,ENDTPA_MIR_Detection,PhysRevApplied.14.064035}. There is enhancement in both the absorption itself, as well as in the current injection that can arise if the appropriate shorter wavelength light is also present to allow for the interfering 1PA. In the ``extremely" non degenerate regimes considered, one photon involved in the two-photon transition is very close to the band gap, while the other can be a MIR photon. This allows for the absorption of infrared photons by large band-gap semiconductors \textit{in the presence} of near band gap photons. Experiments showing the enhancement of ND-2PA using MIR photons over D-2PA have already been performed for GaAs, GaN, ZnSe, and Si \cite{ENDTPA_MIR_Detection,ND2PA_GaAs}.

\begin{figure}[ht]
    \centering
    \includegraphics[width=0.93\linewidth]{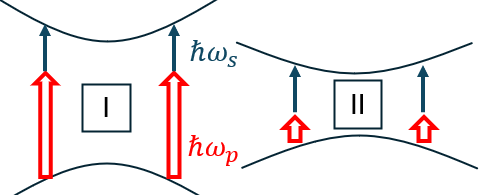}
    \caption{Schematic representation of the ND-2PA process for the same signal photon but different pump and band gap selections. The bold red arrows represent the pump photon energy, and the narrow blue arrows the signal photon energy.}
    \label{fig.ND_2PA_Schematic1}
\end{figure}

To describe ND-2PA, with or without the presence of interfering 1PA, it is helpful to think of the absorption of a signal photon with energy $\hbar\omega_s$ in the presence of a pump photon with energy $\hbar\omega_p$. There are then two schemes that can be applied to MIR detection: (I) the pump photons have energy greater than the signal, or (II) the pump photons have energy less than the signal; see Fig. \ref{fig.ND_2PA_Schematic1}.
These schemes have distinct advantages and disadvantages.

The first scheme is suitable for semiconductors with gaps on the order of 1 eV or larger; for the signal photon in the MIR range a pump laser can be chosen with photon energy close to the band gap. One drawback of this method is that degenerate two-photon absorption of the pump must also be taken into account.

The second scheme is suitable for more narrow gap semiconductors, with band gaps at the edge of the MIR range. Then a pump with photons below half the band gap can increase the sensitivity to signal photons with energies near the band gap. In this scheme there is no D-2PA of the pump with which to contend.

These two schemes can also be applied to the coherent generation of current, the measurement of which would indicate the presence of the signal. Much as there is an enhancement of ND-2PA over D-2PA, there is also an enhancement of the current injection in a three color scheme with beams at $\omega_1$, $\omega_2$, and $\omega_1+\omega_2$, vs a two color scheme with beams at $\omega$ and $2\omega$, where $2\omega=\omega_1+\omega_2$. The most ``extreme" enhancement occurs when one of the photon energies, say $\hbar\omega_1$, approaches the band gap energy. Three-color coherent generation of current has been shown experimentally in GaAs \cite{ThreeColorCoherentControl}, but the ``extremely" non-degenerate regime has not been considered in detail. 

In the search for devices based on one-photon absorption that can operate at non-cryogenic temperatures, the alloy Ge$_{1-x}$Sn$_x$ has been shown to have potential for MIR photon detection \cite{GeSn_OpticalInjection}, since the band structure is tunable \textit{via} strain and composition. Incorporation of $\alpha-$Sn shifts the band structure from an indirect gap semiconductor (pure Ge) to direct gap-semiconductor, with the band gap closing at the $\Gamma$-point at a Sn concentration of approximately $x=0.29$ \cite{Song_2019}. Additionally, the material has been shown to be compatible with silicon processing \cite{SiGeSn_Platform}. The strength of its D-2PA and two-color injected current tensor have been theoretically predicted \cite{GeSn_OpticalInjection}, but the ND-2PA and three-color coherent current generation of Ge$_{1-x}$Sn$_{x}$ has yet to be studied. It would be a natural candidate for the second scheme outlined above.

In this manuscript we calculate the ND-2PA and three color current injection applied to MIR detection for the two schemes: (I) As a reference we consider using GaAs and a near infrared (NIR) pump laser with wavelengths between 0.85 to 0.95 $\mu$m; and (II) we consider using the alloy Ge$_{1-x}$Sn$_{x}$ with a CO$_2$ or quantum cascade laser (QCL) at $\lambda = 10.6 \mu$m as the pump. Our aim is to provide a preliminary characterization of the alloy Ge$_{1-x}$Sn$_x$ for MIR detection using the scheme II discussed above.

Based on well established scaling rules for 2PA \cite{Wherrett,BandGapDependence_2PA,DispersionBandGapScaling,BandGapDependence_ND2PA}, mainly the $E_\text{gap}^{-3}$ dependence, the much smaller band gap of Ge$_{1-x}$Sn$_{x}$ compared to GaAs should indeed prove beneficial in scheme II. However, the scaling rules were derived based on a two-parabolic band approximation, and Ge$_{1-x}$Sn$_{x}$ has contributions to absorption from regions of the Brillouin zone away from the $\Gamma$ point, an anisotropic non-linear response, and significant non-parabolicity as the Sn concentration is increased. Therefore, we employ 30-band $k\cdot p$ models that allow for efficient calculation of the band energies and matrix elements out to the Brillouin zone (BZ) edge. With the 30-band model we provide quantitative predictions for the ND-2PA and three color injected current tensors over a range of Ge$_{1-x}$Sn$_x$ alloy compositions. 

The manuscript is structured as follows. In section \ref{Sec2} we present the theoretical model used to compute the (both D- and ND-) 2PA and the tensor that describes current injection, neglecting many body effects. We also introduce the 30-band $k\cdot p$ models for GaAs and Ge$_{1-x}$Sn$_x$ that allow for integration over the entire BZ. In section \ref{Sec3} we present calculations for the two materials and compare the two schemes introduced above. We also fit our 2PA results to an approximate two-parabolic band formula for quick and efficient exploration of deviations from the frequency pairs we investigated. The results are summarized in section \ref{Sec4}.

\vspace{-10pt}
\section{Theory}\label{Sec2}
\vspace{-5pt}
Many D- and ND-2PA experiments use a pump-probe setup with pulses of light instead of continuous wave (CW) sources \cite{ND_Gain_GaAs,NDTPA_ZincBlende,ND2PA_ExperimentZnX,ENDTPA_MIR_Detection}. This can be helpful in achieving desired peak intensities while not depositing too much energy into the material \cite{ThreeColorCoherentControl}. However, we will take the CW-limit to derive our absorption coefficients, which is standard practice for making theoretical predictions for the magnitude of the absorption coefficients \cite{NDTPA_ZincBlende,ENDTPA_MIR_Detection}. 

We restrict ourselves to the independent particle approximation and consider systems subject to a Hamiltonian $\hat{H} = \hat{H}_0 + \hat{H}_\text{e-em}$, where $\hat{H}_0$ is the Hamiltonian of the crystal in the absence of an external electric field, and $\hat{H}_\text{e-em}$ describes the interaction of the electronic degrees of freedom and the electromagnetic field. The Hamiltonian $\hat{H}_0$ captures the periodic potential of the lattice $V(\textbf{r})$, the kinetic energy, spin-orbit coupling, and correlation effects only in so far as they can be included in an effective lattice potential. The eigenstates of $\hat{H}_0$ are Bloch functions of the form $\psi_{n\textbf{k}}(\textbf{r}) = (2\pi)^{-3/2} e^{i\textbf{k}\cdot\textbf{r}}u_{n\textbf{k}}(\textbf{r})$, the products of plane waves and lattice periodic functions $u_{n\textbf{k}}(\textbf{r})$, with energy eigenvalue $\hbar\omega_{n\textbf{k}}$, labelled by a band index $n$ and crystal momentum $\hbar\textbf{k}$ \cite{CohenCondensedMatter}.

For the expected exciton binding energies of the semiconductors we consider 
\cite{PRB_GaAs_Exciton,PhysRevB_DirectGapExciton,SemiconductorsCardonna},
the independent particle approximation we employ is justified  for the pump and signal configurations we consider. For GaAs the exciton binding energy was measured to be 4.2$\pm0.3$ meV \cite{PRB_GaAs_Exciton}, with the Wannier-Mott excitons theoretically predicted to be $5.7$ meV \cite{PhysRevB_DirectGapExciton}. The theoretical exciton binding energy depends on the reduced effective mass of the heavy-hole and conduction bands, and the relative permittivity of the material. The conduction band effective mass of Ge is smaller than that of GaAs and is made smaller still with Sn incorporation. Additionally, the Ge$_{1-x}$Sn$_{x}$ alloys we consider have a larger dielectric constant than GaAs, leading to stronger screening of the Coulomb interaction. Both the smaller reduced effective mass of the electron-hole pair and larger dielectric constant in Ge$_{1-x}$Sn$_{x}$ alloys leads to a smaller expected exciton binding energy than the 4.2 meV measured in GaAs. In the following we consider absorption at energies away from the exciton peak near the band edge, and off-resonant from the exciton binding energy.

We work in the SI system of electromagnetic quantities and units \cite{JacksonEM} and choose to work in the velocity gauge \footnote{With the use of sum rules the results from a length gauge or velocity gauge are equivalent due to gauge invariance.} to represent the interaction Hamiltonian \cite{VelocityGaugeNLO,VelocityGaugeAnalysis}, 
\begin{equation}
\begin{split}
    \hat{H}_\text{e-em}(\textbf{r},t) = -e A^i(t) \hat{v}^i(\textbf{r}),
\end{split}
\end{equation}
where $e=-|e|$ is the electronic charge, $\hat{v}^i(\textbf{r})$ is the velocity operator, and $\textbf{A}(t)$ is the time-dependent vector potential that describes the electric field, $\textbf{E}(t) = -\partial \textbf{A}(t)/{\partial t}$, with
\begin{equation}
\label{E_Fourier}
    E^i(t) = \sum_{\omega>0} \Big( e^{-i\omega t} E^i(\omega)  + e^{i\omega t} E^{i}(-\omega) \Big).
\end{equation}

\subsection{1- and 2-PA}

From first and second order time-dependent perturbation theory we can obtain a Fermi's golden rule expression for the rate of excited carriers \cite{OpticalInjectionSipe}. The excitation of carriers from the valence band to the conduction band by one photon absorption (1PA) for monochromatic sources is described by the equation
\begin{equation}
\begin{split}
    \dot{n}^{(1)} = \xi^{(1)}_{ab}(\omega) E^a(-\omega) E^b(\omega),
\end{split}
\end{equation}
with \cite{OpticalInjectionSipe}
\begin{equation}
\label{Xi1}
\begin{split}
    \xi^{(1)}_{ab}(\omega) = \frac{2\pi e^2}{\hbar^2 \omega^2} \int_{BZ} \frac{d\textbf{k}}{(2\pi)^3} v^a_{vc} v^b_{cv} \delta(\omega_{cv}-\omega),
\end{split}
\end{equation}
and where $c$ labels conduction bands, $v$ labels valence bands, and $v^a_{nm}$ are the velocity matrix elements in the eigenstate basis of $\hat{H}_{0}$. Here $\dot{n}$ is the rate of change of the number density of injected carriers, and has units of m$^{-3}$s$^{-1}$. We neglect any indirect absorption processes mediated by phonons here and in the other processes we consider below. The tensor $\xi^{(1)}_{ab}(\omega)$ is related to the imaginary part of the linear electric susceptibility $\chi_{ab}(\omega)$, $\text{Im}\Big[ \chi_{ab}(\omega) \Big] = \hbar \xi^{(1)}_{ab}(\omega)/{2\epsilon_0}$.

Results are often quoted under the assumption that the relative permittivity tensor at frequencies of interest, $\epsilon^{r}_{ab}(\omega) = \delta_{ab} + \text{Re}\Big[\chi_{ab}(\omega)\Big]$, is a multiple of the unit tensor, with $\epsilon^r_{xx}(\omega) = \epsilon^r_{yy}(\omega) = \epsilon^r_{zz}(\omega) \equiv \epsilon^r(\omega)$
\cite{NDTPA_ZincBlende,ENDTPA_MIR_Detection,ND_Gain_GaAs}. This is true of both GaAs and Ge$_{1-x}$Sn$_x$; the index of refraction $n(\omega) = \sqrt{\epsilon^r(\omega)}$, and for a monochromatic plane wave with a time dependence of the form of eq. (\ref{E_Fourier}), linearly polarized along the $x$ direction $\hat{e}_x$ and propagating along the $z$ direction, the intensity is given by $I(\omega) = 2\epsilon_0 n(\omega) c |E^x({\omega})|^2$. 
We generally adopt this scenario below, in which case the tensor components on the absorption coefficients are all $x$. But more generally, for two-photon absorption and a chosen propagation direction and polarization, the relevant effective absorption coefficient will be a superposition of different Cartesian components, and can be easily worked out.

As the beam propagates through the material the intensity as a function of $z$ (the direction of propagation) is described by the equation
\begin{equation}
    \frac{dI(\omega)}{dz} = -\alpha^{(1)}(\omega) I(\omega), 
\end{equation}
where $\alpha^{(1)}(\omega)$ is the 1PA absorption coefficient. Within the above approximations for the state of the electric field the absorption coefficient can be calculated from the imaginary part of the electric susceptibility $\alpha^{(1)}_{xx}(\omega) = \hbar\omega \xi^{(1)}_{xx}(\omega)/{2\epsilon_0 n(\omega) c}$ \cite{OpticalPropertiesRolfBinder}. This relation is obtained by noting that for every electron injected into a conduction band a photon of energy $\hbar\omega$ is absorbed, and thus the rate of injected carriers can be expressed in terms of the 1PA absorption coefficient, 
\begin{equation}
\label{n1_I}
\begin{split}
    \dot{n}^{(1)} = \frac{\alpha^{(1)}(\omega)}{\hbar\omega} I({\omega}).
\end{split}
\end{equation} 

In a two beam configuration with neither beam having frequencies compatible with  1PA, but where 2PA can exist, the rate of injected carriers is given by
\begin{equation}
\label{ndot_2}
\begin{split}
    \dot{n}^{(2)} = \xi^{(2)}_{abcd}(\omega_1,\omega_2) E^a(-\omega_1) E^b(-\omega_2) E^c(\omega_1) E^d(\omega_2).
\end{split}
\end{equation}
The tensor is calculated by evaluating a single integral over the BZ \cite{OpticalInjectionSipe}
\begin{equation}
\label{Xi2_Tensor}
\begin{split}
    \xi^{(2)}_{abcd}(\omega_1,\omega_2) &= 2\pi \sum_{cv} \int_{BZ} \frac{d\textbf{k}}{(2\pi)^3} \Big(\Gamma^{ab}_{cv\textbf{k}}(\omega_1,\omega_2) \Big)^*
    \\
    &\hspace{20pt}\times \Gamma^{cd}_{cv\textbf{k}}(\omega_1,\omega_2) \delta(\omega_{cv}-\omega_1-\omega_2),
\end{split}
\end{equation}
with
\begin{equation}
\label{Gamma_cvk}
\begin{split}
    \Gamma^{ab}_{cv\textbf{k}}(\omega_1,\omega_2) = \frac{e^2}{2\hbar \omega_1 \hbar\omega_2} \Bigg[ 
    &\sum_{c'} \Big[ \frac{v^b_{cc'}v^a_{c'v}}{\omega_{c'v} - \omega_1} + \frac{v^a_{cc'}v^b_{c'v}}{\omega_{c'v}-\omega_2} \Big]
    \\
    - &\sum_{v'} \Big[ \frac{v^a_{cv'}v^b_{v'v}}{\omega_{cv'}-\omega_1} + \frac{v^b_{cv'}v^a_{v'v}}{\omega_{cv'}-\omega_2} \Big]
    \Bigg].
\end{split}
\end{equation}
Due to the Dirac delta function in eq. (\ref{Xi2_Tensor}) we can implicitly take $\Gamma^{ab}_{cv\textbf{k}}(\omega_1,\omega_2)\rightarrow \Gamma^{ab}_{cv\textbf{k}}(\omega_1,\omega_{cv}-\omega_1)$, simplifying the frequency dependence of the function.

A more commonly quoted quantity is the two-photon absorption coefficient $\alpha_{2}(\omega_1;\omega_2)$. This describes the absorption of light at frequency $\omega_1$ in the presence of light at frequency $\omega_2$. Indicating the intensity of light at $\omega_1$ by $I_1$, and the intensity of light at $\omega_2$ by $I_2$, we have
\begin{equation}
\label{IntensityAbsorption}
\begin{split}
    \frac{\partial I_1}{\partial z} = -\alpha^{(2)}(\omega_1;\omega_1) I^2_1 -2 \alpha^{(2)}(\omega_1;\omega_2) I_1 I_2 ,
    \\
    \frac{\partial I_2}{\partial z} = -\alpha^{(2)}(\omega_2;\omega_2) I^{2}_{2} - 2\alpha^{(2)}(\omega_2;\omega_1) I_{1} I_{2}.
\end{split}
\end{equation}
The D-2PA is described by $\alpha^{(2)}(\omega_1;\omega_1)$ and $\alpha^{(2)}(\omega_2;\omega_2)$, while ND-2PA is described by $\alpha^{(2)}(\omega_1;\omega_2)$ and $\alpha^{(2)}(\omega_2;\omega_1)$. There is also a relation involving the interchange of the order of the frequencies in the 2PA absorption coefficient:  
\begin{equation}\label{ND2PA_Relation}
    \alpha^{(2)}(\omega_1;\omega_2) = \frac{\omega_1}{\omega_2} \alpha^{(2)}(\omega_2;\omega_1).
\end{equation}
Thus, for $\omega_1 > \omega_2$ we have $\alpha^{(2)}(\omega_1;\omega_2) > \alpha^{(2)}(\omega_2;\omega_1)$, as can be understood physically: For each photon with energy $\hbar\omega_1$ absorbed, there is also a photon with energy $\hbar\omega_2$ absorbed, and whichever photon individually has more energy necessarily deposits more energy to the material. 

To determine an expression for $\alpha^{(2)}$ we proceed as was done in obtaining eq. (\ref{n1_I}), and identify
\begin{equation}
\label{n2_absorption}
\begin{split}
    \dot{n}^{(2)} = \alpha^{(2)}(\omega_1;\omega_1)\frac{I_1^2}{2\hbar\omega_1} &+ \alpha^{(2)}(\omega_2;\omega_2) \frac{I_2^2}{2\hbar\omega_2} 
    \\
    &+2\alpha^{(2)}(\omega_1;\omega_2) \frac{I_1 I_2}{\hbar\omega_1},
\end{split}
\end{equation}
leading to \cite{ND2PA_QuantumWells}
\begin{equation}
\label{alpha2}
    \alpha^{(2)}_{abcd}(\omega_1;\omega_2) = \frac{\hbar\omega_1}{2\epsilon_0^2 n_1 n_2 c^2} \xi^{(2)}_{abcd}(\omega_1;\omega_2), 
\end{equation}
where we have introduced the subscript notation for the index of refraction evaluated at a particular frequency, $n_1 = n(\omega_1)$ and $n_2 = n(\omega_2)$. 
If both fields at $\omega_1$ and $\omega_2$ are polarized in the $x$-direction and propagating along the $z$-direction we would only require the $\alpha^{(2)}_{xxxx}$ component. For different configurations of the fields at $\omega_1$ and $\omega_2$, different components of the tensor $\alpha^{(2)}$ will be required to estimate the absorption. 

By examining eq. (\ref{Xi2_Tensor}) and eq. (\ref{Gamma_cvk}) it is clear that the 2PA is enhanced if the two frequencies $\omega_1$ and $\omega_2$ are very different. For example, if $\omega_1$ is much smaller than $\omega_2$, then $\Gamma^{ab}_{cv\textbf{k}}$ is roughly proportional to $\omega_1^{-1}$, and $\xi^{(2)}$ is roughly proportional to $\omega_1^{-2}$.  

\vspace{-10pt}
\subsection{Current injection}
\vspace{-5pt}
For coherent current injection we consider bi- or tri-chromatic fields described by eq. (\ref{E_Fourier}) where the set of frequencies are $\omega_1$, $\omega_2$, and $\omega_1+\omega_2$. The bichromatic field ``limit" can be taken by setting $\omega_1=\omega_2=\omega$ and $E_{\omega_1} = E_{\omega_2} = \frac{1}{2}E_{\omega}$. Taking the limit in this way allows both D- and ND-2PA to be characterized by the tensor defined in eq. (\ref{alpha2}). 

The charge current has an injection rate given by \cite{OpticalInjectionSipe,GeSn_OpticalInjection,ThreeColorCoherentControl}
\begin{equation}
\begin{split}
    \dot{J}_{a}^{I} = \eta_{abcd}^{I}(\omega_1,\omega_2) E^{b*}(\omega_1) E^{c*}(\omega_2) E^d(\omega_1+\omega_2) + c.c.,
\end{split}
\end{equation}
with the fourth rank tensor $\eta^{I}$:
\begin{equation}
\label{eta_eh}
\begin{split}
    &\eta_{abcd}^{I ;\substack{e \\ (h)}}(\omega_1,\omega_2) = \substack{+\\(-)} \frac{i\pi e^2}{\hbar(\omega_1+\omega_2)} \int_{BZ} \frac{d\textbf{k}}{(2\pi)^3} \sum_{cv} 
    \\
    &\hspace{50pt} \Bigg[ 
    \sideset{}{'}\sum_{\substack{c' \\ (v') }} \Big( \Gamma^{bc}_{cv\textbf{k}}(\omega_1,\omega_2)\Big)^* v^a_{\substack{cc' \\ (v'v)}} v^d_{\substack{c'v \\ (cv')}} 
    \Bigg] 
    \\
    &\hspace{50pt} \Bigg[ 
    \delta(\omega_{\substack{c'v \\ (cv')}}-\omega_1-\omega_2)
    + \delta(\omega_{cv}-\omega_1-\omega_2)
    \Bigg],
\end{split}
\end{equation}
where there is both an electron contribution $\eta_{abcd}^{I;e}$ and hole contribution $\eta_{abcd}^{I;h}$. The prime notation on the sum and band indices indicates that at the given $\mathbf{k}$ the sum over $c'(v')$ is a sum over band $c(v)$ and any other bands with the same energy as band $c(v)$ \cite{PhysRevB.76.205113,PhysRevB.81.155215}. We have used a stacked notation $\substack{e\\(h)}$ to write both contributions in eq. (\ref{eta_eh}); the top indicates the electron contribution and the bottom the hole contribution.  

Since $\eta^{I}$ depends only linearly on the matrix elements $\Gamma^{bc}_{cv\textbf{k}}$, whereas $\xi^{(2)}$ depends quadratically on them, the enhancement in the extremely non-degenerate limit is expected to be less substantial for current injection than it is for two-photon absorption, but still present.

\subsection{Thirty-band model : GaAs}

\begin{figure}
\centering
\includegraphics[width=0.88\linewidth]{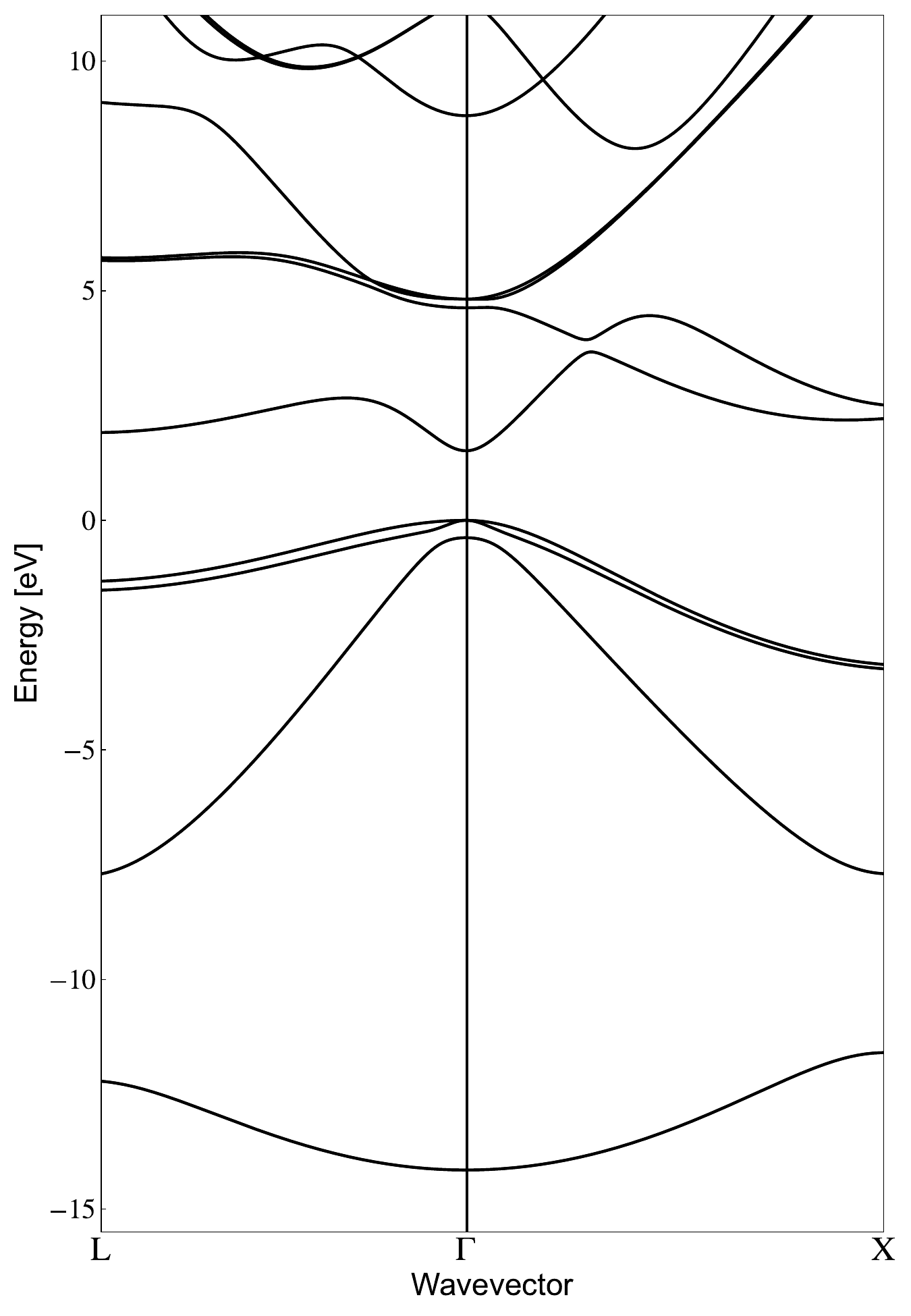}
\caption{GaAs band structure along two high symmetry directions in the BZ, calculated with the 30-band $k\cdot p$ model.}
\label{fig:GaAsBands}
\end{figure}

We employ a 30-band $k\cdot p$ model for determining the band structure of GaAs across the entire BZ. The model was derived for general III-V compounds, and so the 30 basis states are grouped according to the irreducible representations of the $T_d$ point group ($\Gamma_6$, $\Gamma_8$, or $\Gamma_7$). The 24 parameters we use for modeling GaAs are from Gawarecki et al. \cite{GaAs_Model}, and are listed in Table \ref{tab:GaAs} of the Appendix. The explicit form of the 30$\times$30 Hamiltonian can be found in Appendix A of Gawarecki et al. \cite{GaAs_Model}. 

Fig. \ref{fig:GaAsBands} shows the GaAs band structure along two high symmetry directions, which is in good agreement with density functional theory (DFT) results for the six highest valence bands and eight lowest conduction bands (see Fig. 2 in Gawarecki et al. for the comparison of the $k\cdot p$ model parameters we use to DFT \cite{GaAs_Model}).

\begin{figure*}
    \centering
    \subfloat[Evolution of the direct and indirect band gap with increasing Sn concentration, as calculated with the 30-band $k\cdot p$ model.]{%
        \label{fig:GeSnx_Gap}%
        \includegraphics[width=0.45\textwidth]{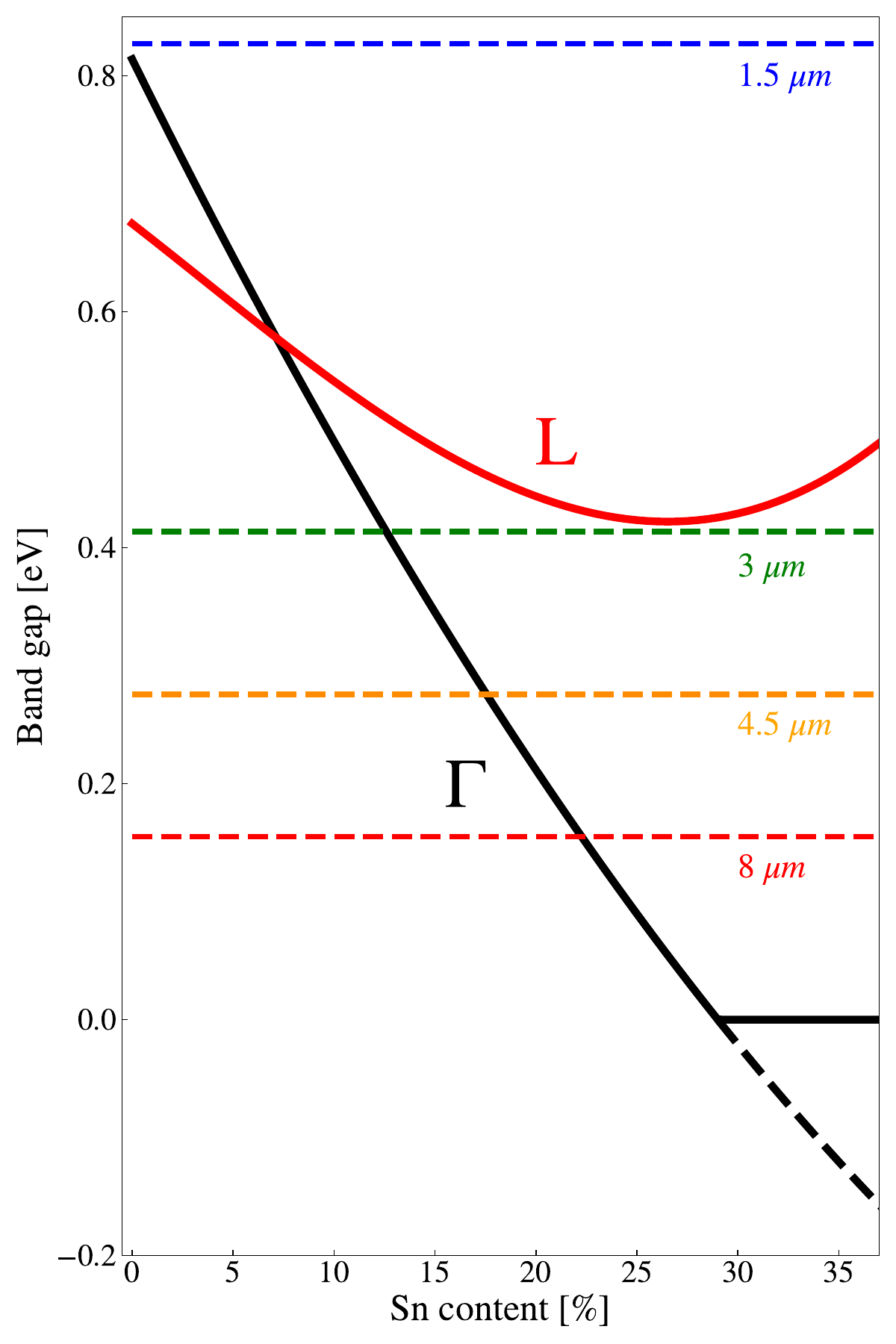}%
    }%
    \hfill 
    \subfloat[Ge and Ge$_{0.83}$Sn$_{0.17}$ band structure along two high symmetry directions in the BZ, calculated with the 30-band $k\cdot p$ model.]{%
        \label{fig:GeSnxBands}%
        \includegraphics[width=0.45\textwidth]{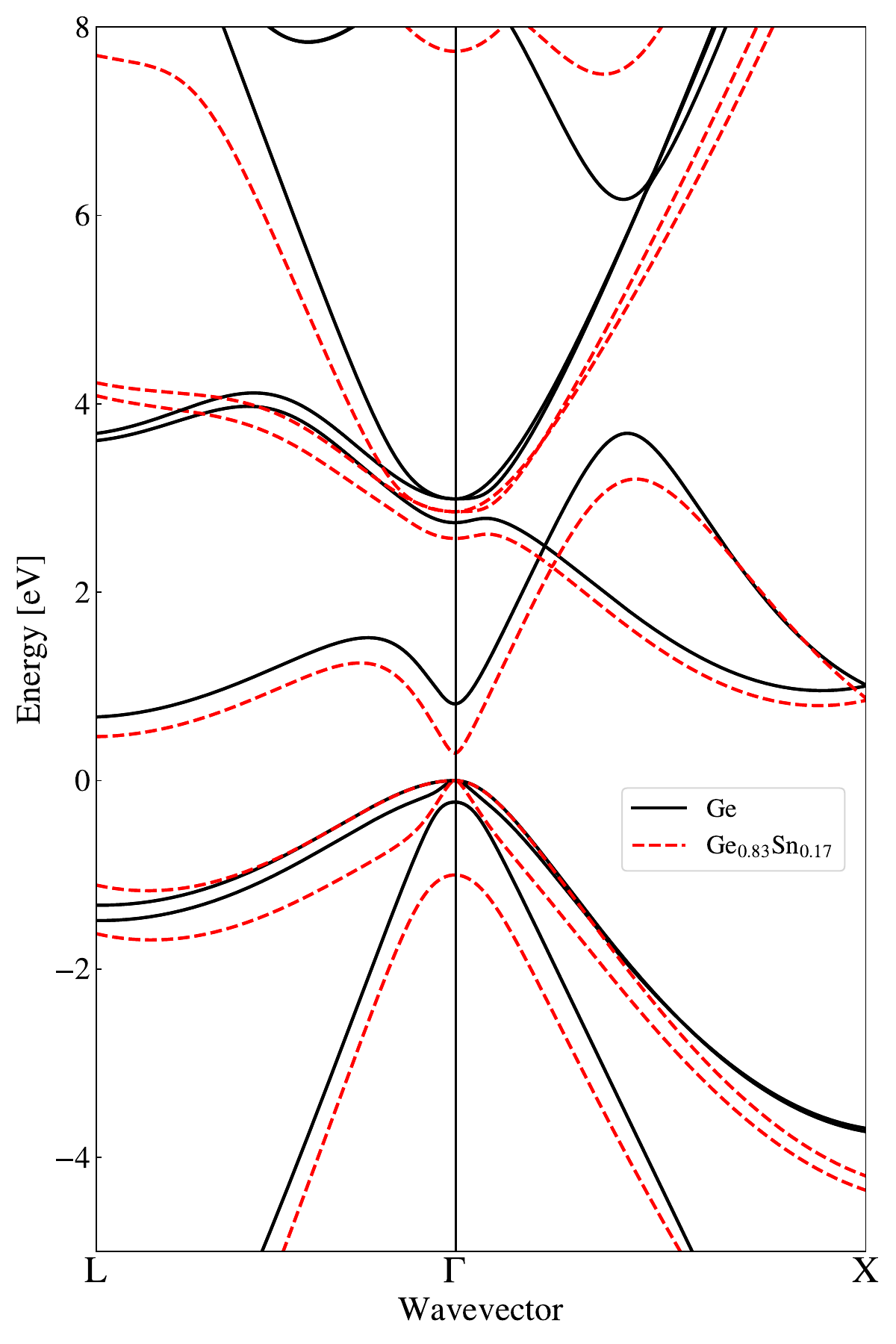}%
    }%
    \caption{Dependence of Ge$_{1-x}$Sn$_x$ band structure on Sn concentration.}
    \label{fig:placeholder}
\end{figure*}

\subsection{Thirty-band model : \texorpdfstring{Ge$_{1-x}$Sn$_{x}$}{GeSn}}

In Ge-based semiconductors an accurate description of the optical response requires capturing features around the band edges, such as the L point. For this reason we use a 30-band model that reproduces the band structure across the entire BZ at energies near the band gap \cite{Song_2019}. 

The incorporation of $\alpha-$Sn is handled in the virtual crystal approximation (VCA) with up to quadratic interpolation fitted with available experimental data. This model predicts the transition from indirect to direct band gap at $x=7.2\%$, and the transition to a semi-metal above $x=29\%$. Fig. \ref{fig:GeSnx_Gap} shows the evolution of the band gap at the L and $\Gamma$ point as the Sn concentration is increased.

At $x=29\%$ the four highest valence bands intersect with the lowest pair of conduction bands. The pair of conduction bands and one of the pairs of valence bands have a linear dispersion near the $\Gamma$ point, while the second pair of valence bands has a quadratic dispersion. For $x>29\%$ the linear bands go back to a quadratic dispersion, and the formerly linear valence band pair pulls away from the other bands. The pair of valence bands that remains quadratic at $x=29\%$ remains degenerate with the pair of conduction bands. This band shuffling is indicated in Fig. \ref{fig:GeSnx_Gap}: The dashed black line tracks the pulling away of one pair of valence bands, leading to a lower energy, while the solid black line shows the persistent degeneracy of the other pair of valence bands with the conduction bands. Additionally, there is a band inversion that occurs at $x=29\%$; the gap closing and band inversion indicate that Ge$_{1-x}$Sn$_{x}$ could host a topological Dirac semimetal phase \cite{TopologicalGeSn}. But this is just at the edge of validity of the model, and the development of more accurate models to study this transition is warranted; it is out of the scope of this work.

The model of Song et al. \cite{Song_2019} is optimized using experimental data that spans the range of $x \in [0,0.2]$ \cite{PRB_Exp_1,FittingExperiment_2,PRB_Exp_3,FittingExperiment_4}. Beyond this range, near the semi-metal transition that we do not consider, local structural disorder or alloy clustering may cause results to deviate from the virtual crystal approximation. The highest concentration we consider in this work is $x=20\%$, which is at the edge of the experimental data used for the quadratic fitting of the model parameters, so we believe will provide quantitatively reliable results.

We assume the crystal belongs to the $O_h$ group, ignoring any local inversion symmetry breaking in the alloy, and use the room temperature parameters from Song et al. \cite{Song_2019}. The 22 parameters are listed in Table \ref{tab:GeSn}, and the explicit form of the Hamiltonian can be found in section 2 of Song et al \cite{Song_2019}. Fig. \ref{fig:GeSnxBands} shows the band structure of Ge and Ge$_{0.83}$Sn$_{0.17}$ calculated using the 30-band $k\cdot p$ model along two high symmetry directions. There is significant lowering of the split-off valence, as well as lowering of the conduction band  with Sn incorporation; the latter is most pronounced at the $\Gamma$ point. The gradual shift of  the lowest pair of conduction bands and a pair of valence bands from a quadratic to linear dispersion near the $\Gamma$ point can also be seen. 

\subsection{Computational Details}

The integral over the BZ is performed by employing a linear tetrahedron method \cite{TetrahedronMethod}. The tensors are only evaluated on a discrete set of points restricted to the irreducible wedge of the BZ, with either the $T_d$ or $O_h$ group operations used to produce the result for integration over the entire BZ. To improve computation speeds we group the calculation into batches of $k$ points and distribute them to separate processor cores for simultaneous calculations.

The delta functions in eqs. (\ref{Xi1}), (\ref{Xi2_Tensor}), and (\ref{eta_eh}) are approximated by a Gaussian
\begin{equation}
    \delta(\omega_{cv}-\omega) \rightarrow \frac{\exp( -\frac{1}{2} \frac{(\omega_{cv}-\omega)^2}{\gamma^2})}{\gamma\sqrt{2\pi}}.
\end{equation}
We choose a width $\hbar\gamma$ between 5 and 15 meV to obtain smooth results for the tensors as a function of frequency, while still preserving features of the absorption coefficients. As we scan over photon energies, we stop our calculations when the signal photon energy is sufficient to resonantly connect a valence band with a conduction band.  At this photon energy and higher, a more general theory of two-photon absorption that takes into account resonant excitation of an intermediate state would be necessary. 

All code is an ``in-house" implementation of the 30-band $k\cdot p$ models, using a BZ integration scheme for the linear and nonlinear response tensors; it is available upon request.

\section{Two-Schemes Comparison} \label{Sec3}

In the non-degenerate regime of 2PA we distinguish the two photons that interact within the material, labelling one $\omega_s$ for ``signal" and the other $\omega_p$ for ``pump;" the pump photons are supplied by a strong beam of light, whereas light is to be detected at the signal frequency. In both scheme I and scheme II only one of the signal and pump photons has energy greater than half the band gap, and ideally both have energy less than the band gap, \textit{i.e.,} there is no 1PA. For a current injection experiment, a third photon with energy $\hbar\omega_s+\hbar\omega_p$ that leads to the interfering 1PA process must also be included. 

\begin{figure}[ht]
    \centering
    \includegraphics[width=0.7\linewidth]{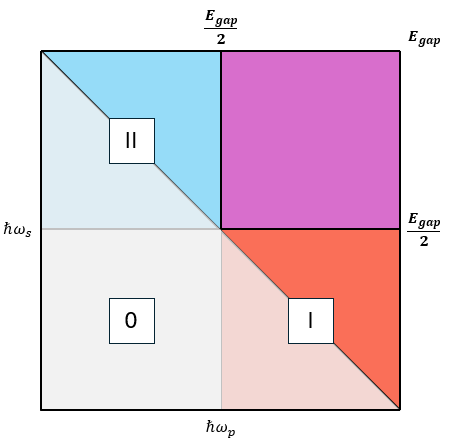}
    \caption{Representation of where in the parameter space of signal and pump frequencies relative to the material band gap scheme I and scheme II exist. In the lower off-diagonal shaded in grey $\hbar\omega_p+\hbar\omega_s < E_\text{gap}$ so no 2PA is possible. } 
    \label{fig:SchemeRegions}
\end{figure}

Fig. \ref{fig:SchemeRegions} shows a schematic representation of these signal and pump frequency choices. In region 0, encompassing the entire bottom left quadrant and extending to the off diagonal in region I and II, there is no 2PA possible within our approximations, since $\hbar\omega_p+\hbar\omega_s < E_\text{gap}$. In region I the pump photon has energy greater than half the band gap, and ND-2PA is possible only when $\hbar\omega_s + \hbar\omega_p > E_\text{gap}$. In region II the signal photon has energy greater than half the band gap, with the same energy constraint. The top right quadrant includes D-2PA as well as ND-2PA with both energies greater than half the band gap.  

While D-2PA depends on the square of the intensity of the light at the signal frequency, ND-2PA depends on the product of the intensities of the signal and pump beams. This offers an additional opportunity to enhance the absorption, since one can separately control the intensity of the pump beam. 

On experimentally measuring the effectiveness of these schemes, three possibilities can be considered: transmission and reflection of either the signal or pump beam; photoluminescence \cite{MultiphotonAbsorptionProperties}; or applying a DC bias voltage to convert the injected carriers into a current. In scheme I, the pump photon will lead to D-2PA as well as facilitate ND-2PA of the signal. With a strong pump beam intensity the carrier injection due to the signal photons may be ``washed" out by the injected carriers due to the D-2PA of the pump, complicating the measurement. In scheme II, there is no D-2PA of the pump, and the signal photon will undergo both D-2PA and assisted ND-2PA. Due to the enhancement of ND over D-2PA, and the scaling of ND-2PA with the pump intensity, the ND-2PA should dominate. Furthermore, neglecting the effects of defects and impurities, the material will be transparent to the pump beam. This is ideal since any injected carriers will require the absorption of a signal photon. Interestingly, a measurement of the pump transmission would also be an implicit measurement of the signal absorption, since only in the presence of the signal photons will there be absorption of the pump beam.

Without applying a DC field current can still be generated by the interference of the 1- and 2-PA processes. The coherent generation of current can then be characterized either by a current measurement or voltage accumulation in a device \cite{ThreeColorCoherentControl}. We use the same material and pump photon pairs as above, but also include a third electromagnetic field with photon energies $\hbar\omega_p+\hbar\omega_s$ for the coherent generation of current calculations.

For scheme I we employ a GaAs model with band gap of 1.514 eV, and four possible pump frequencies at [1.30, 1.35, 1.40, 1.45] eV, corresponding to vacuum wavelengths of [0.954, 0.918, 0.886, 0.855] $\mu$m. For scheme II we employ a Ge$_{1-x}$Sn$_{x}$ model with band gap at the $\Gamma$ point shown in Fig. \ref{fig:GeSnx_Gap}, and a pump at 0.117 eV, corresponding to a vacuum wavelength of 10.6 $\mu$m. To prevent D-2PA of the pump\textcolor{blue}{,} we cannot exceed a Sn concentration of 20\%.

\begin{figure}[h]
    \centering
    \includegraphics[width=0.73\linewidth]{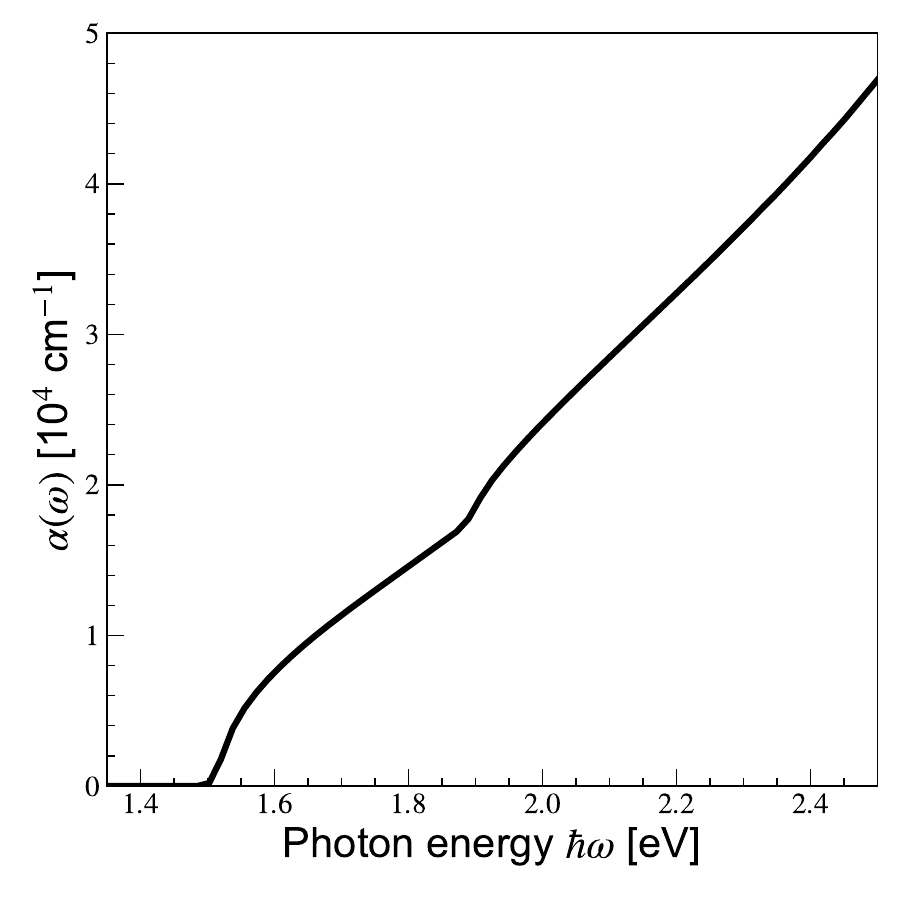}
    \caption{1PA of GaAs as a function of incident photon energy $\hbar\omega$ near the band gap, calculated with the 30-band $k\cdot p$ model.}
    \label{fig:1PA_GaAs}
\end{figure}

\begin{figure*}[ht]
    \centering
    \subfloat[ND-2PA of GaAs as a function of signal photon energy $\hbar\omega_s$ in the presence of four different pump energies $\hbar\omega_p$, calculated with the 30-band $k\cdot p$ model. Only the largest component $\alpha^{(2)}_{xxxx}$ is shown.]{%
        \label{fig:ND2PA_GaAs}%
        \includegraphics[width=0.48\textwidth]{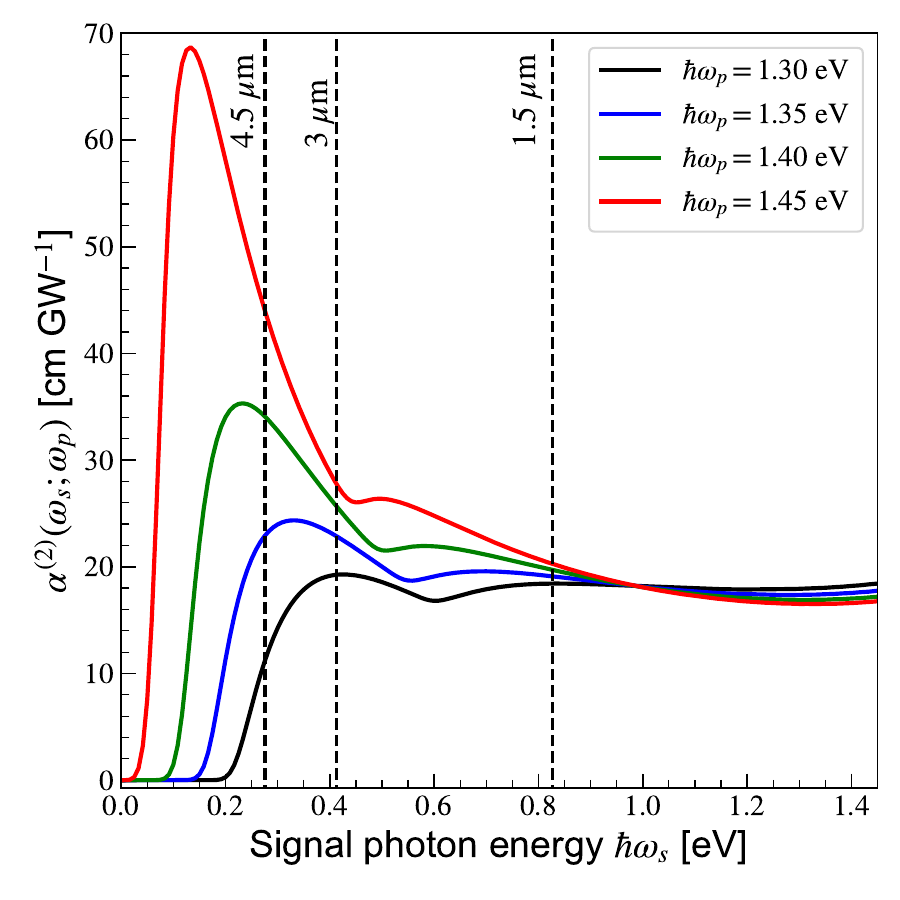}%
    }%
    \hfill 
    \subfloat[The three independent components of the D-2PA tensor of GaAs as a function of incident photon energy $\hbar\omega$, calculated with the 30-band $k\cdot p$ model.]{%
        \label{fig:D2PA_GaAs}%
        \includegraphics[width=0.48\textwidth]{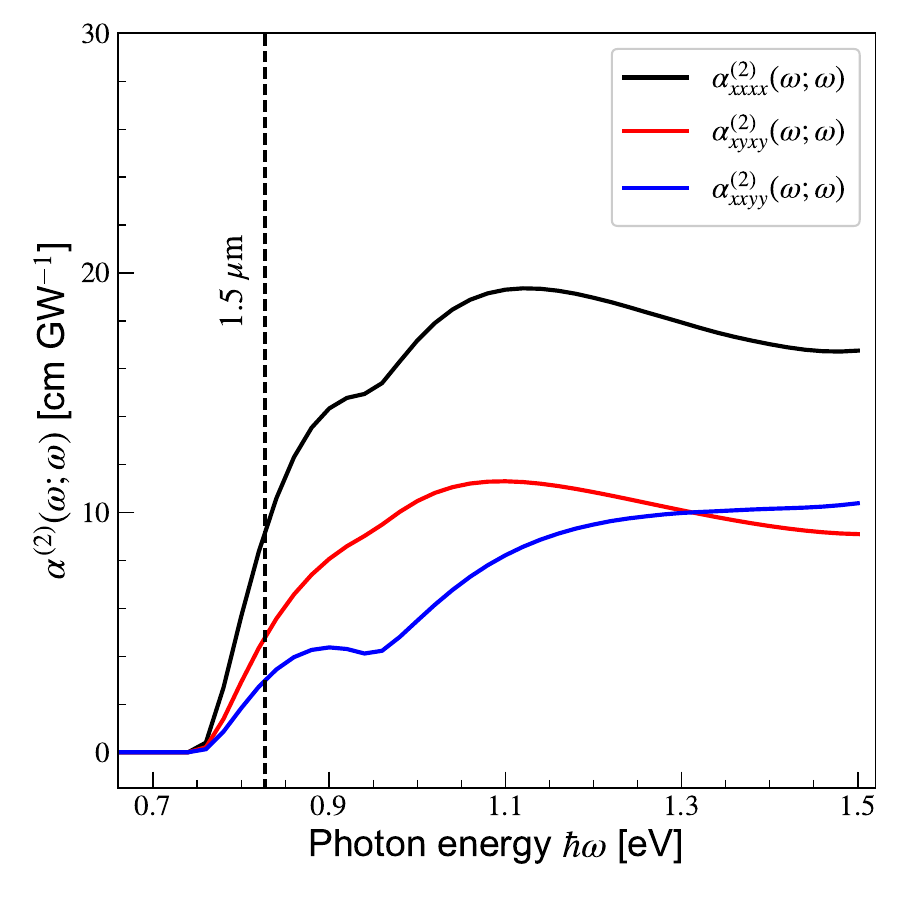}%
    }%
    \caption{ND- and D-2PA of GaAs over a range of pump and signal photon energies.}
    \label{fig:placeholder2}
\end{figure*}

\subsection{Scheme I - GaAs}

The 1PA of GaAs near the band gap is adequately described by the transitions near the $\Gamma$ point. Within 1 eV of the band gap the absorption is on the order of $10^4$ cm$^{-1}$, which we will use as a kind of ``figure of merit" for assessing the feasibility of the two schemes for MIR 2PA. In Fig. \ref{fig:1PA_GaAs} we plot only the $\alpha^{(1)}_{xx}$ component as a function of frequency since the 1PA tensor is isotropic for GaAs.   

Due to the tetrahedral symmetry of GaAs there are four independent components to the ND-2PA tensor; $\alpha^{(2)}_{xxxx}(\omega_s;\omega_p)$, $\alpha^{(2)}_{xyxy}(\omega_s;\omega_p)$, $\alpha^{(2)}_{xyyx}(\omega_s;\omega_p)$, and $\alpha^{(2)}_{xxyy}(\omega_s;\omega_p)$. The other tensor components are obtained by performing the following operations on $\alpha^{(2)}_{abcd}(\omega_s;\omega_p)$: simultaneously exchanging the order of the first pair of indices and the second pair of indices $a \leftrightarrow b, c\leftrightarrow d$; exchanging the first and second pair of indices $ab \leftrightarrow cd$; or performing a permutation of Cartesian indices; all tensor components not reachable in this way are zero. In total there are 21 non-zero tensor components for this crystal class \cite{OpticalTensors}.

We plot the $\alpha^{(2)}_{xxxx}(\omega_s;\omega_p)$ component of the ND-2PA absorption tensor for a range of pump photon energies in Figure \ref{fig:ND2PA_GaAs}. As the pump energy approaches the band gap, the absorption spectrum is pulled to lower energies and its maximum value grows. Thus, in the presence of a NIR or visible pump photon GaAs can absorb photons with wavelengths longer than 4.5 $\mu$m. For NIR pumps with intensities of 1 TW cm$^{-2}$ bulk GaAs has a theoretical absorption of MIR photons comparable to its natural 1PA near the band gap. But such a high pump intensity is not required for an observable effect, since 1PA is non-existent in GaAs at these low photon energies. In fact, non-degenerate 2PA pump probe experiments have been successfully performed with peak pulse irradiances outside the sample between 0.7 to 9 GW cm$^{-2}$ \cite{ND2PA_ExperimentZnX,ND_Gain_GaAs,ENDTPA_MIR_Detection}. For one of the GaAs experiments, with $\hbar\omega_p = 1.27$ eV and $\hbar\omega_s$=0.16 eV, absorption coefficients of $\alpha^{(2)}_{||} =(38 \pm 8)$ cm/GW and $\alpha^{(2)}_{\perp}=(16\pm 4)$ cm/GW for parallel and perpendicular polarization configurations were measured \cite{ND_Gain_GaAs}. Although the band structure of Gawarecki et al. \cite{GaAs_Model} that we use takes the band gap to be its value at zero temperature, from Figure \ref{fig:ND2PA_GaAs} it is clear that our results are generally consistent with those experimental measurements. We must use different pump and signal frequencies due to the different band gap of our model, but as an example we find for $\hbar\omega_p = 1.40$ eV and $\hbar\omega_s$=0.175 eV, which has a similar non-degeneracy ratio to the experiment, $\alpha^{(2)}_{||}=30.2$ cm/GW and $\alpha^{(2)}_{\perp} = 19.5$ cm/GW. The smaller $\alpha_{||}$ is to be expected due to the larger band gap of our model. Our focus here is not to compare to these experiments directly, but to demonstrate possible improvement over the existing schemes. 

\begin{figure}[ht]
    \centering
    \includegraphics[width=0.85\linewidth]{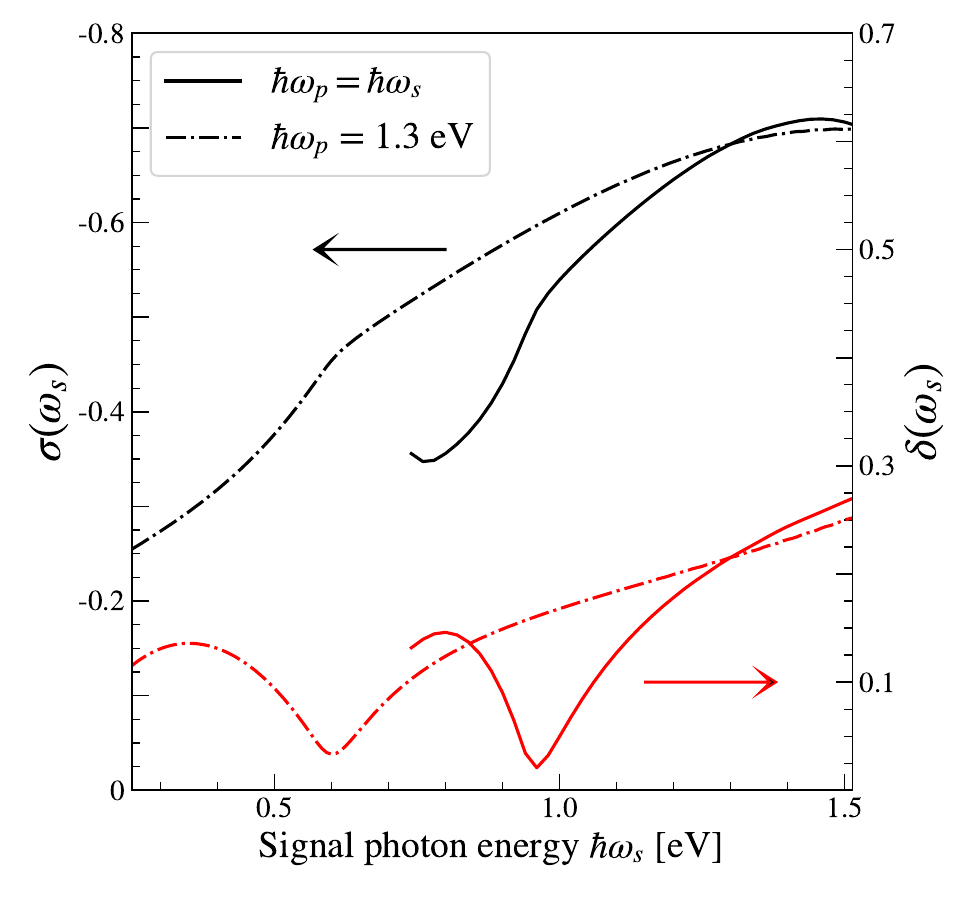}
    \caption{The spectral dependence of the anisotropy and linear-circular dichroism for D-2PA and ND-2PA with a pump at $\hbar\omega_p = 1.3$ eV, calculated with the 30-band $k\cdot p$ GaAs model. $\sigma$ plotted in black uses the left axis, while $\delta$ plotted in red uses the right axis.}
    \label{fig:Anisotropy_GaAs}
\end{figure}

\begin{figure*}
    \centering
    \includegraphics[width=0.92\linewidth]{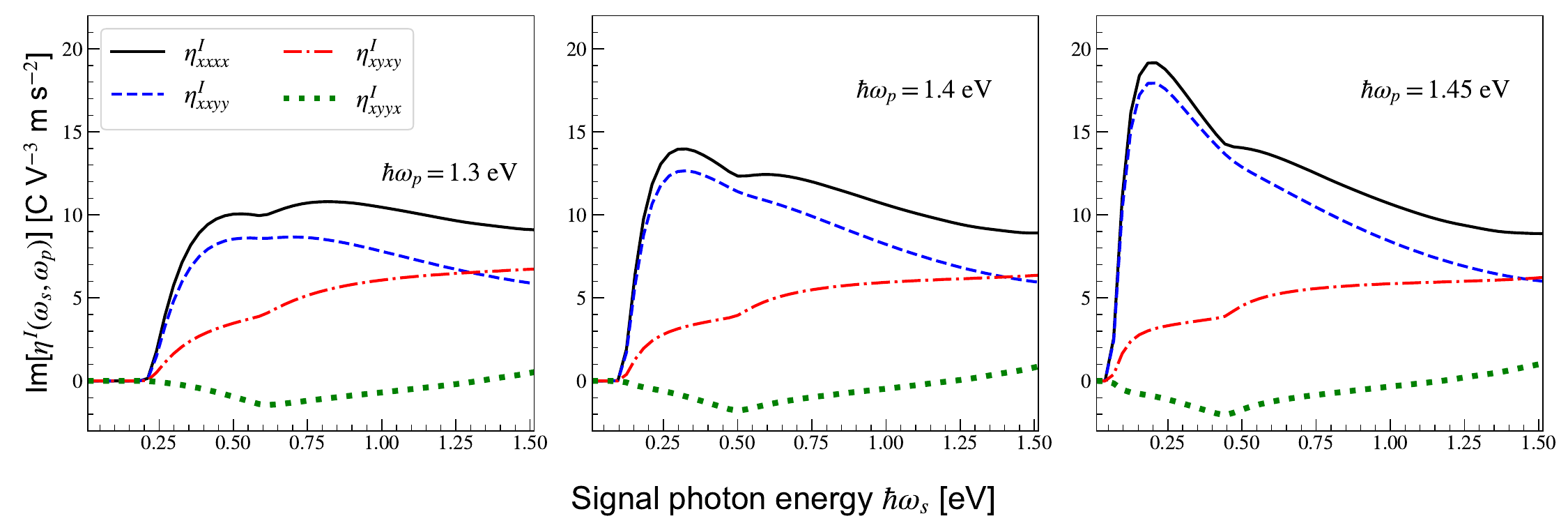}
    \caption{Three-color injected current tensor $\eta^{I}(\omega_{s},\omega_{p},\omega_{s}+\omega_p)$ of GaAs as a function of signal photon energy $\hbar\omega_s$ for three different pump energies $\hbar\omega_p$, calculated with the 30-band $k\cdot p$ model.}
    \label{fig:etaND_GaAs}
\end{figure*}

In scheme I, since the pump photons have energy greater than half the band gap, there will also be D-2PA of the pump but no D-2PA of the signal. Depending on the experimental detection method, i.e. measuring injected carriers, or through transmission of the pump, one must also calculate and account for $\alpha^{(2)}_{abcd}(\omega_p;\omega_p)$. For the D-2PA process there are only three independent tensor components $\alpha^{(2)}_{xxxx}$, $\alpha^{(2)}_{xyxy}$, and $\alpha^{(2)}_{xxyy}$. The other tensor components are obtained by performing any of the following operations on the tensor $\alpha^{(2)}_{abcd}$: exchanging the first two indices $a\leftrightarrow b$; exchanging the second two indices $c \leftrightarrow d$; exchanging the first and second pair of indices $ab \leftrightarrow cd$; or performing a permutation of Cartesian indices ($x \rightarrow y\rightarrow z\rightarrow x$); all tensor components not reachable in this way are zero. Like with the ND-2PA tensor there are a total of 21 non-zero tensor components \cite{OpticalTensors}, however, tensor elements like $\alpha^{(2)}_{xyxy}$ and $\alpha^{(2)}_{xyyx}$ that were distinct are now equal. The three independent tensor components for the D-2PA process are shown in Fig. \ref{fig:D2PA_GaAs}.

The highest pump used in Figure \ref{fig:ND2PA_GaAs} is approximately $96\%$ of the band gap (theoretical model, $E_\text{gap}$ = 1.514 eV), with an absorption peak near $0.13$ eV and a value of $\alpha^{(2)}(0.13 \text{eV},1.45 \text{eV})$ just shy of 70 cm GW$^{-1}$. Based on the ND-2PA relation of eq. (\ref{ND2PA_Relation}) the effective pump absorption would be $\alpha^{(2)}(1.45 \text{eV},0.13 \text{eV}) \approx 770$ cm GW$^{-1}$, and $\alpha^{(2)}(1.45 \text{eV}, 1.45 \text{eV}) \approx 17$ cm GW$^{-1}$. From these calculations we see that the ND-2PA of a 1.45 eV pump laser in the presence of a MIR photon can be up to 45$\times$ the degenerate coefficient. However, if the pump intensity is greater than the MIR intensity the D-2PA is comparatively increased by the ratio of the intensities.

To obtain very quick approximate results for any frequency combinations without needing to perform the BZ integration we fit our results to a two-parabolic band formula \cite{Wherrett,BandGapDependence_2PA,BandGapDependence_ND2PA,DispersionBandGapScaling}:
\begin{equation}
\label{ParabolicBandFormula}
\begin{split}
    \alpha^{(2)}_{xxxx}(\omega_s;\omega_p) = K\frac{(x_s+x_p-1)^{3/2} (x_s+x_p)^2}{2^8 x_s^3 x_p^4},
\end{split}
\end{equation}
where $x_i = \frac{\hbar\omega_i}{E_\text{gap}}$, and $K$ is a material dependent constant. For the GaAs model we employ $K_\text{GaAs} \approx 850$ cm GW$^{-1}$.   

For general configurations of the electric field there can be anisotropy of the 2PA if one rotates the plane of polarization for linearly polarized light, as well as a difference in absorption of circular vs linear polarized light. To succinctly represent these properties one can introduce three scalar quantities \cite{OpticalInjectionSipe,Anisotropy1,Anisotropy2}: the strength $\xi^{(2)}_{xxxx}$; the anisotropy parameter $\sigma$
\begin{equation}
    \sigma = \frac{\xi^{(2)}_{xxxx}-(\xi^{(2)}_{xyxy}+\xi^{(2)}_{xyyx}+\xi^{(2)}_{xxyy})}{\xi^{(2)}_{xxxx}};
\end{equation}
and the linear-circular dichroism $\delta$
\begin{equation}
\begin{split}
    \delta = \frac{\xi^{(2)}_{xxxx} + \xi^{(2)}_{xxyy} - (\xi^{(2)}_{xyxy}+\xi^{(2)}_{xyyx})}{2\xi^{(2)}_{xxxx}}.
\end{split}
\end{equation}

A value of $\sigma=0$ indicates an isotropic model with no dependence on crystal orientation. A value of $\delta' = \delta-\sigma/6 = 0$ indicates that the 2PA is identical for linear and circular polarized light \cite{OpticalInjectionSipe}. 

The spectra of $\sigma$ and $\delta$ obtained from the 30-band $k\cdot p$ model are presented in Fig. \ref{fig:Anisotropy_GaAs}. The anisotropy grows until the signal-frequency nears the band gap energy where it reaches a peak of approximately $\sigma = -0.7$. The linear-circular dichroism is non-zero over the frequencies we consider. As we sweep the photon energy $\delta$ initially drops to a minimum then tends to grow with increasing signal frequency energy. The magnitude and shape of the ND-2PA anisotropy and linear-circular dichroism are very similar to that of the D-2PA. However, they are stretched out over the frequency range since the pump pushes the onset of absorption to lower energies.   

We then consider adding a third incident beam with photons of energy $\hbar\omega_{s}+\hbar\omega_p$ that leads to 1PA that can interfere with the ND-2PA process. If the relative phases of the beams are adjusted this can lead to an injected current. For zinc-blende crystal structures there are four independent components to describe the 3-color injected current tensor: $\eta^{I}_{xxxx}$, $\eta^{I}_{xxyy}$, $\eta^{I}_{xyyx}$, and $\eta^{I}_{xyxy}$. In total there are 21 non-zero tensor components $\eta^{I}_{abcd}$ obtained from these by \cite{OpticalTensors}: (1) cyclic permutations of the indices; (2) simultaneous exchange of $b \leftrightarrow c$ and $a \leftrightarrow d$; and (3) $ab \leftrightarrow cd$. The results for the spectral dependence of the total electron+hole contributions to the four independent tensor components of the 3-color coherent control are presented in Fig. \ref{fig:etaND_GaAs} for a range of pumps. As the pump photon energy approaches the energy gap the spectrum is naturally pulled to lower signal photon energies. We also see the $\eta^{I}_{xxxx}$ and $\eta^{I}_{xxyy}$ components grow while the other tensor components do not grow appreciably with higher pump photon energies. 

Experimental results for three-color beams of energies 0.86, 0.69, and 1.55 eV had a maximum steady-state voltage of approximately 1-3 mV for an irradiance between 10-15 MW cm$^{-2}$ \cite{ThreeColorCoherentControl}.

These results for GaAs under scheme I are 
used as a comparison for the Ge$_{1-x}$Sn$_{x}$ results that follow in the next subsection.

\subsection{\texorpdfstring{Scheme II - Ge$_{1-x}$Sn$_{x}$}{Scheme II - GeSn}}

\begin{figure}[h!]
    \centering
    \includegraphics[width=0.75\linewidth]{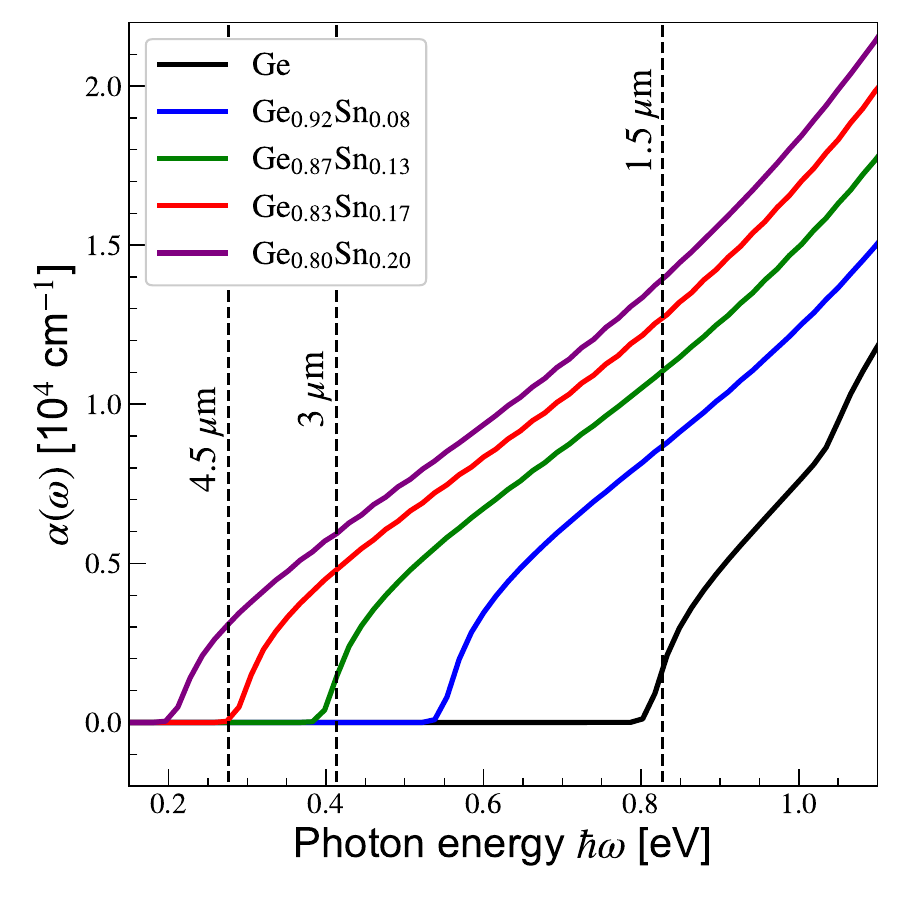}
    \caption{1PA of Ge, Ge$_{0.92}$Sn$_{0.08}$, Ge$_{0.87}$Sn$_{0.13}$, Ge$_{0.83}$Sn$_{0.17}$, and Ge$_{0.80}$Sn$_{0.20}$, as a function of incident photon energy $\hbar\omega$ near the band gap, calculated with the 30-band $k\cdot p$ model.}
    \label{fig.1PA_GeSn}
\end{figure}

Due to the octahedral symmetry we assume for the Ge$_{1-x}$Sn$_{x}$ crystal, the 1PA tensor has only one independent component. The shift of the onset of the 1PA to longer wavelengths with increasing Sn content is evident in Fig \ref{fig.1PA_GeSn}. The onset of direct gap absorption for pure Ge is close to 1.5 $\mu$m, for Ge$_{0.87}$Sn$_{0.13}$ close to 3 $\mu m$, and for Ge$_{0.83}$Sn$_{0.17}$ just below 4.5 $\mu m$. The order of magnitude is similar to that of GaAs, but at much lower photon energies. For the Ge crystal structure the D-2PA, ND-2PA, and three-color coherent control tensors follow the same symmetry rules as the GaAs tensors explained above.

\begin{figure*}
    \centering
    \includegraphics[width=0.9\linewidth]{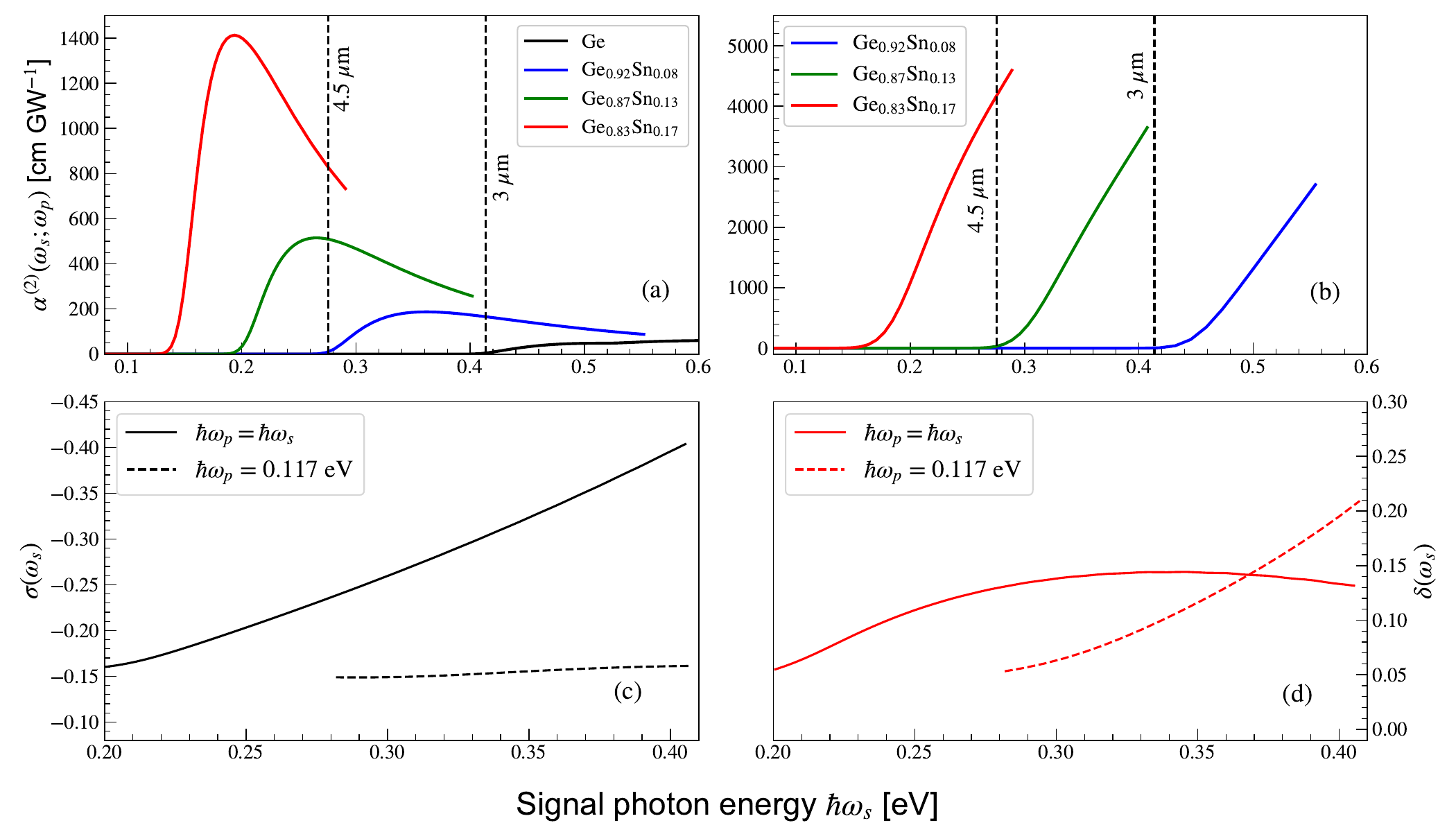}
    \caption{(a) The $\alpha^{(2)}_{xxxx}$ component of the D-2PA tensor for four Ge$_{1-x}$Sn$_{x}$ alloys, calculated with the 30-band $k\cdot p$ model. (b) The $\alpha^{(2)}_{xxxx}$ component of the ND-2PA tensor as a function of signal frequency $\hbar\omega_s$ for a pump energy $\hbar\omega_p = 0.117$ eV. (c) [(d)] Spectral dependence of the anisotropy (linear-circular dichroism) of the D-2PA and ND-2PA ($\hbar\omega_p=0.117$ eV) tensors for the alloy Ge$_{0.87}$Sn$_{0.13}$. All plots are terminated when the signal photon energy reaches that particular alloy's bandgap. }
    \label{fig:GeSnx_alphaDND}
\end{figure*}

In scheme II there will be D-2PA of the signal, but not of the pump. Thus any absorption of the pump, ignoring material defects and impurities, is associated with the absorption of photons at the signal frequency. We choose a pump photon energy of $0.117$ eV corresponding to a vacuum wavelength of 10.6 $\mu$m. For this pump photon energy we are in the regime of scheme II ($\hbar\omega_p < E_\text{gap}/2$) for concentrations $x<20$\%.

We only show the largest component of the D-2PA and ND-2PA tensor components ($\alpha^{(2)}_{xxxx}$) for a range of Ge$_{1-x}$Sn$_x$ alloys in Fig. \ref{fig:GeSnx_alphaDND} (a) and (b) respectively. As expected, the onset of 2PA occurs at lower energies with higher Sn concentration and the peak magnitude also increases. The peak of the degenerate absorption coefficient for Ge$_{0.83}$Sn$_{0.17}$ is 20$\times$ the peak of the ND-2PA of GaAs in a similar signal photon energy range, cf. Fig. \ref{fig:ND2PA_GaAs} and Fig. \ref{fig:GeSnx_alphaDND} (a). 

One disadvantage of scheme II is that the onset of ND-2PA occurs at higher signal photon energies relative to D-2PA\textcolor{blue}{,} due to the pump being less than half the band gap. However, in the range of frequencies that are allowed we see a significant enhancement in the absorption coefficient. For example, at $x=17\%$ we find $\alpha^{(2)}_{xxxx}$(0.275 \text{eV}, 0.117 \text{eV}) $\approx 4100$ cm GW$^{-1}$. We now only require a pump intensity of 1 GW cm$^{-2}$ to obtain an effective absorption coefficient for $4.5 \mu$m light that is 4100 cm$^{-1}$. Peak pump irradiances on the order of 1 GW cm$^{-2}$ have been used in pump probe experiments \cite{ND_Gain_GaAs,ND2PA_ExperimentZnX} with 2PA absorption coefficients two orders of magnitude smaller, so we believe there is experimental potential for Ge$_{1-x}$Sn$_{x}$ with this ND-2PA scheme.   

To quickly consider other pairs of signal and pump photons, we need only use eq. (\ref{ParabolicBandFormula}) with approximate material dependent constants $K_\text{Ge} \approx 2700$, $K_\text{Ge$_{0.92}$Sn$_{0.08}$} \approx 7300$, $K_\text{Ge$_{0.87}$Sn$_{0.13}$} \approx 20300$, and $K_\text{Ge$_{0.83}$Sn$_{0.17}$} \approx 56400$, all in cm GW$^{-1}$. The fit to the two-parabolic band formula becomes worse as we go to higher Sn concentrations and towards more non-degenerate photon pairs, possibly due to the increased non-parabolicity of the bands with higher Sn concentration.  

We also evaluate the anisotropy and linear circular-dichroism for Ge$_{0.87}$Sn$_{0.13}$, for both the D- and ND-2PA tensors, presented in Fig. \ref{fig:GeSnx_alphaDND} (c) and (d). The ND-2PA anisotropy is effectively flattened over the frequency range where 1PA is not-allowed\textcolor{blue}{,} and has a smaller absolute value compared to the D-2PA anisotropy. In contrast, the linear-circular dichroism of the ND-2PA rises more sharply than that of the D-2PA.   

\begin{figure}[h!]
    \centering
    \includegraphics[width=0.75\linewidth]{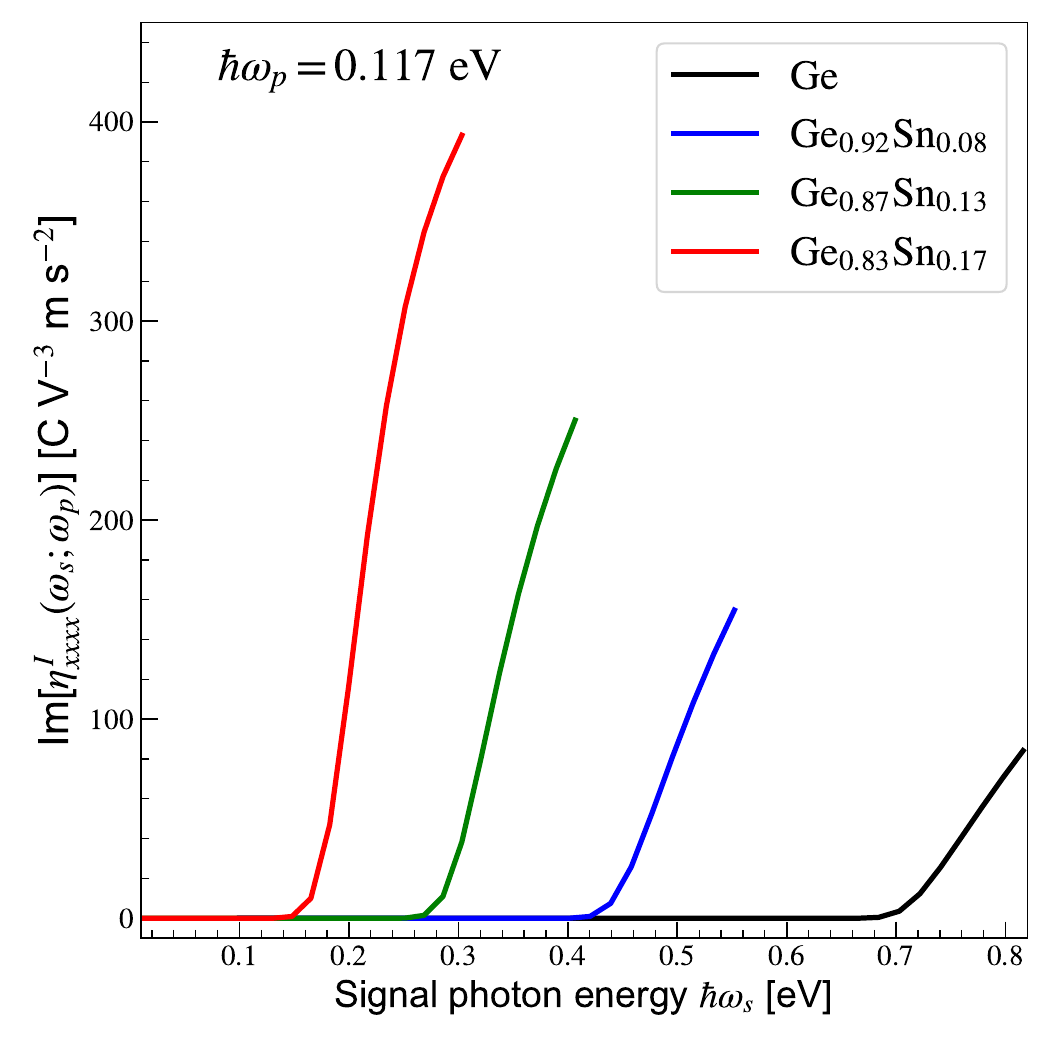}
    \caption{3-color injected current tensor (sum of electron and hole contributions) for four different Ge$_{1-x}$Sn$_x$ alloys. Plots are terminated when the signal photon energy reaches that particular alloy's bandgap.}
    \label{fig:etaND_GeSnx}
\end{figure}

The spectrum of the 3-color injected current tensor $\eta^{I}_{xxxx}$ is presented in Fig. \ref{fig:etaND_GeSnx}. The magnitude of this tensor component is much larger than that of GaAs in regions where both materials have the interference of 1- and 2PA, cf. Fig. \ref{fig:etaND_GaAs}. The three-color scheme does not experience as large an increase over the two-color scheme as does D-2PA vs ND-2PA, albeit an enhancement is present. Comparing values in the MIR region, the three-color GaAs results are between 10-20 C V$^{-3}$ m s$^{-2}$, whereas for the Ge$_{1-x}$Sn$_{x}$ alloys we consider the results are between 100-400 C V$^{-3}$ m s$^{-2}$.

The absorption and injected current coefficients obtained above for Ge$_{1-x}$Sn$_{x}$ are larger than values already measured experimentally \cite{ThreeColorCoherentControl,ND2PA_ExperimentZnX,ND_Gain_GaAs} in what we have dubbed ``scheme I." We believe this indicates there is practical feasibility with available lasers and material processing. 

While we have compared bulk absorption coefficients in the preceding sections, as a detection figure of merit the responsivity at a given frequency, $R(\omega_s)$, which measures the ratio of photocurrent to incident optical power, may be a better figure of merit. A rudimentary expression for the responsivity can be obtained by first considering an effective focusing area $\mathcal{A}$ of the incident beam at $\omega_s$. The optical power would then be $P_s = I_{s}\mathcal{A}$. Using eq. (\ref{IntensityAbsorption}), we can estimate the rate of energy absorbed from the field as it propagates through a material of thickness $d$. The resulting rate of injected carriers, $\dot{N}_e$, is then given by 
\begin{equation}
    \dot{N}_e = \frac{P'_s}{\hbar\omega_s}(1-e^{-2\alpha^{(2)}(\omega_s;\omega_p)I'_p d}),
\end{equation}
where $I'_p$ is the pump intensity \textit{inside} the material and $P_{s}'$ is the optical power of the signal \textit{inside} the material. We have assumed the pump intensity is much greater than the signal intensity, so we omit the degenerate 2PA process. The pump and signal power inside the material will be reduced due to reflection at the surface of the semiconductor interface. Therefore, assuming a reflection coefficient of $R_f$, we must amend the power in the material by a factor of $(1-R_f)$. There will be additional losses like from recombination that we include into a factor $\zeta$. Thus, the output photocurrent can be approximated as
\begin{equation}
    I_{e} = e\zeta \frac{P_s}{\hbar\omega_s}(1-R_f)(1-e^{-2\alpha^{(2)}(\omega_s;\omega_p)I'_p d}).
\end{equation}
Dividing by the input optical power we obtain an expression for the responsivity at $\omega_s$
\begin{equation}
\label{Responsivity}
    R(\omega_s) \approx \zeta(1-R_f) (1-e^{-2\alpha^{(2)}(\omega_s;\omega_p) I'_p d}) \frac{\lambda_s}{1.24},
\end{equation}
where the wavelength of the signal light is $\lambda_s$ microns \cite{sze_physics_2007,kasap2013optoelectronics}. In a device with active absorption region $d$ there may also be reflection at the back interface depending on the materials used. This is neglected in eq. (\ref{Responsivity}), where only a single pass of the signal and pump beams is considered.  

From eq. (\ref{Responsivity}) we see how the detector responsivity is enhanced not only by a larger nonlinear coefficient $\alpha^{(2)}$, but also by increasing the pump power $I_p$ and the detector thickness $d$. Due to the larger index of refraction of the Ge$_{1-x}$Sn$_{x}$ alloys, mitigating losses due to reflection should be considered in device fabrication. Additionally, the requirement that the pulse and signal beams overlap in the device must be addressed to benefit from the enhanced non-degenerate absorption coefficient.

Assuming a detector thickness of 5 $\mu$m, a reflectivity $R_f=0.5$, a pump intensity of 1 GW cm$^{-2}$ at 0.117 $\text{eV}$, neglecting recombination losses, and with an absorption coefficient of $\alpha^{(2)} \approx 4100$ cm GW$^{-1}$ for signal photons at 4.5 $\mu$m (0.275 eV), the responsivity is $R(0.275 \text{ eV}) \approx1.78$ A W$^{-1}$. While this is a simple calculation with many assumptions, it exceeds existing measured responsivities in GaN detectors \cite{ENDTPA_MIR_Detection}, so we consider further investigation is warranted.

\section{Conclusion}\label{Sec4}

We have compared two schemes for MIR detection using non-degenerate 2PA: (I) using a material with a band gap in the NIR or optical range then applying a strong pump near the band gap to allow for MIR absorption; or (II) using a semiconductor with a gap already in the MIR range, but enhancing MIR detection below the band gap by applying a strong pump at energies below half the band gap. In addition, either of these schemes can then be used to inject current when a third beam with photon energies of $\hbar\omega_p+\hbar\omega_s$ is included. 

To quantitatively compare schemes I and II we performed theoretical calculations for GaAs with pump photon energies between 1.3 and 1.45 eV (I), and for Ge$_{1-x}$Sn$_x$ alloys with a pump photon energy of 0.117 eV/$\lambda$ = 10.6 $\mu$m (II). To include the effects of anisotropy and non-parabolicity of the band structure we have used a 30-band $k\cdot p$ model for the calculations.

We found that for a range of compositions the material Ge$_{1-x}$Sn$_{x}$ has enhanced sensitivity to MIR photons with wavelengths above $1.5$ $\mu$m to even larger than 4.5 $\mu$m. This sensitivity is present in both the 2PA and the injected current response due to the interference of 1PA and 2PA. The nonlinear coefficients we find are significantly larger in the Ge$_{1-x}$Sn$_{x}$ alloys we consider than in the wider gap material GaAs employing scheme I. For example, we find a ND-2PA coefficient of $\alpha^{(2)}_{xxxx}$(0.275 eV, 0.117 eV)$\approx$ 4100 cm GW$^{-1}$ for Ge$_{0.83}$Sn$_{0.17}$, which we believe constitutes a reasonable choice of pump photon and an optimistic but possible alloy composition. Comparatively, for GaAs we found $\alpha^{(2)}_{xxxx}$(0.275 eV, 1.45 eV)$\approx$ 45 cm GW$^{-1}$ and $\alpha^{(2)}_{xxxx}$(1.45 eV, 0.275 eV)$\approx$ 240 cm GW$^{-1}$. However, the maximum of the GaAs response is at a lower energy, 
$\alpha^{(2)}_{xxxx}(0.13 \text{eV},1.45 \text{eV}) \approx 70$ cm GW$^{-1}$ and $\alpha^{(2)}_{xxxx}(1.45 \text{eV},0.13 \text{eV}) \approx 770$ cm GW$^{-1}$.

We also find that the theoretically predicted three-color coherent generation of current in Ge$_{1-x}$Sn$_{x}$ is substantially larger than that in GaAs, with values of the injected current coefficient $\eta^{I}_{\text{Ge}_{0.83}\text{Sn}_{0.17}} \approx 400$ and $\eta^{I}_{\text{GaAs}}\approx 20$ C V$^{-3}$ m s$^{-2}$ for the pump wavelengths we chose. Of course, if pump photon energies closer to the gap can be employed in scheme I, or lower energy pump photons in scheme II, while avoiding the linear Urbach tail absorption or the exciton tail absorption, these results can be improved.    

In general, scheme I and II have their own advantages and disadvantages. Scheme I ($\hbar\omega_p > E_\text{gap}/2)$ allows for the absorption of very low energy signal photons since the pump photon energies can be taken close to the gap, and the remaining energy to cross the gap is provided by the signal photon. However, the D-2PA of the pump has to be accounted for in addition to the desired ND-2PA of the MIR photon. This limits how strong the pump beam can be. 

In contrast to scheme I, in scheme II there is no D-2PA of the pump with which to contend. While the onset of ND-2PA of the signal occurs at higher photon energies than the D-2PA of the signal, and thus both ND-2PA and D-2PA of the signal will occur, for strong pump pulses the ND-2PA should dominate. To achieve signal sensitivity in the MIR frequency range the band gap of the material must be less than 1 eV. The alloy Ge$_{1-x}$Sn$_{x}$, with a direct band gap less than 0.814 eV, satisfies this latter condition. And pump photons with energy less than half the band gap are possible due to the existence of CO$_2$ or quantum cascade lasers for alloy concentrations up to x=20\%.

We hope these promising numbers for the initial characterization of the ND-2PA and three-color injected current using scheme II and the alloy Ge$_{1-x}$Sn$_x$ promote further investigation into actual devices based either on this alloy or absorption scheme. The absorption and injected current coefficients here are larger than those already observed experimentally using ``scheme I." And although more detailed calculations and device designs will be necessary before any practical applications can be envisioned, we believe there is potential in exploring more generally the parameter space available to us with these ND-2PA schemes. In particular, (1) more than just optical or NIR lasers can be considered as pumps; as we have shown, lasers that can operate at 10 $\mu$m or longer can be used to enhance sensitivity at the intermediate wavelengths between $1.5-10$ $\mu$m; (2) different pairings of available lasers and materials to tune to the ``extremely" non-degenerate regime could be employed; (3) band gap engineering with alloying and stress and strain \cite{Song_2019} can be used to expand the catalogue of useful materials; and (4) exploring these effects in heterostructures, such as quantum wells \cite{rotaru2025holespindirectbandgap}, will likely be advantageous for ``scheme II" scenarios as it has already been shown to be for ``scheme I" configurations \cite{ND2PA_QuantumWells}.

\section{Acknowledgements}

The authors acknowledge support from NSERC Canada and Defence Canada (Innovation for Defence Excellence and Security, IDEaS). We also thank O. Moutanabbir and N. Rotaru for helpful discussions on theory and fabrication of the Ge$_{1-x}$Sn$_{x}$ alloy.

\newpage 

\appendix 

\begin{widetext}
\section*{30-band \texorpdfstring{$k \cdot p$}{k.p} model parameters}\label{AppendixTables}

The matrix elements used in the 30-band $k\cdot p$ models are presented in Tables \ref{tab:GaAs} and \ref{tab:GeSn}.

\begin{table}[ht]
    \centering
    \caption{GaAs parameters for the 30-band $k\cdot p$ model}
    \begin{ruledtabular}
    \begin{tabular}{l c c c c c c c} 
    & (eV) &  & (eV) & & (eV \text{\AA}) & & (eV \text{\AA}) \\ [3pt]
    \colrule
         $E_{1c}$ & 1.514 & $E_{5d}$ & 12.800 & $P_0$ & 9.343 & $Q_0$ & 8.350
         \\
         $E_{1w}$ & -14.149 & $E_{1q}$ & 15.662 & $P_1$ & 0.256 & $Q_1$ & -5.106
         \\
         $E_{5v}$ & -0.126 & $\Delta_{v}$ & 0.378 & $P_2$ & 2.152 & $R_0$ & 4.538
         \\
         $E_{5c}$ & 4.754& $\Delta_{c}$ & 0.191 & $P_3$ & 9.332 & $R_1$ & 6.095
         \\
         $E_{1u}$ & 8.811 & $\Delta_{d}$ & 0.030 & $P_4$ & 8.372 & $P'_0$ & -0.509$i$
         \\
         $E_{3t}$ & 11.267 & $\Delta^{-}$ & -0.038$i$ & $P_5$ & 2.389 & $P'_1$ & 2.455$i$
        \\
    \end{tabular}
    \end{ruledtabular}
    \label{tab:GaAs}
\end{table}

\vspace{10pt}

\begin{table}[ht]
    \centering
    \caption{Ge$_{1-x}$Sn$_{x}$ parameters for the 30-band $k\cdot p$ model}
    \begin{ruledtabular}
    \begin{tabular}{l c c c}
    & (eV) &  & (eV) \\ [3pt]
    \colrule
    $\Gamma_{1^l}$ & -12.2519 + 1.4249$x$& $\Gamma_{1^u}$ & 8.2064 - 2.7334$x$ 
    \\
    $\Gamma_{2'^l}$ & 0.8140 - 3.4667$x$ + 2.2767$x^2$ & $\Gamma_{2'^u}$ & 17.0426 - 5.5226$x$ 
    \\
    $\Gamma_{12'}$ & 8.5786 - 0.9856$x$ & $\Delta_{\Gamma_{25'^l},\Gamma_{25'^u}}$ & 0.22 + 0.336$x$ 
    \\
    $\Gamma_{25'^u}$ & 13.4041 - 4.8581$x$ & $\Delta_{25'^u}$ & 0.0793 - 0.0333$x$
    \\
    $\Gamma_{25'^l}$ & 0.00 &$\Delta_{25'^l}$ & 0.2247 + 5.3808$x$ - 4.9535$x^2$
    \\
    $\Gamma_{15}$ & 2.990 - 0.796$x$ & $\Delta_{15}$ & 0.2520 + 0.193$x$
    \\ [3pt]
    \hline \hline \noalign{\smallskip}
    & (eV nm) & & (eV nm) \\ [3pt]
    \colrule
    $P_1$ & 0.8421 + 0.3350$x$ - 0.3346$x^2$ & $Q_2$ & -0.5334 -1.4875$x$ + 2.4647$x^2$
    \\
    $P_2$ & 0.1781 - 2.1954$x$ + 1.9328$x^2$ & $R_1$ & 0.3757 + 0.0340$x$ + 0.0089$x^2$
    \\
    $P_3$ & -0.0734+1.3200$x$-1.2452$x^2$ & $R_2$ & 0.6820 - 1.1743$x$
    \\
    $P_4$ & 1.0543 + 0.2174$x$ - 0.4220$x^2$ & $T_1$ & 0.7994-0.0565$x$ - 0 .0441$x^2$
    \\
    $Q_1$ & 0.8114 - 0.1332$x$ - 0.0055$x^2$ & $T_2$ & -0.0384 + 0.3675$x$
    \end{tabular}
    \end{ruledtabular}
    \label{tab:GeSn}
\end{table}
\end{widetext}

\FloatBarrier

\bibliography{TwoSchemes.bib}

\end{document}